\documentclass[aps,pra,preprint,superscriptaddress,showpacs,preprintnumbers,amsmath,amssymb]{revtex4-1}
\usepackage{graphicx}

\usepackage{xcolor}
\definecolor{ForestGreen}{RGB}{34,139,34}

\begin{document}

\title{Optimal optomechanical cavity setups with highly reflecting membranes}
\author{Georg Enzian}
\affiliation{Niels Bohr Institute, Quantum Optics Laboratory - QUANTOP, Blegdamsvej 17, DK-2100 Copenhagen, Denmark}
\author{Eugene S. Polzik}
\affiliation{Niels Bohr Institute, Quantum Optics Laboratory - QUANTOP, Blegdamsvej 17, DK-2100 Copenhagen, Denmark}
\author{Alexander K. Tagantsev}
\email{alexander.tagantsev@epfl.ch}
\affiliation{Swiss Federal Institute of Technology (EPFL), School of Engineering, Institute of Materials Science, CH-1015 Lausanne, Switzerland}
\begin{abstract}
Highly reflecting mechanically compliant membranes based on photonic-crystal patterns have recently gained increasing attention within cavity optomechanics due to their prospects of reaching high coupling rates in membrane-in-the-middle experiments.
Here we present an analysis and comparison of four different setups in which highly reflecting membranes can be employed for cavity optomechanics, and discuss optimal choices w.r.t. the figures of merit: cooperativity and efficiency-weighted cooperativity.
The analysis encompasses three different types of membrane-in-the-middle setups (membrane-at-the-edge, membrane-in-the-actual-middle, and membrane-at-the-back), as well as the simple Fabry-Perot cavity.
Interestingly, we identify and propose the  membrane-at-the-back setup as an optimal choice in the limit of negligible membrane parasitic loss, which can reach enormous enhancements of optomechanical cooperativity, and if implemented with a low-loss membrane would pave a way to nonlinear optomechanics in the quantum regime.
 \end{abstract}

\date{\today}

\maketitle

\newpage

\section{Introduction}
\begin{figure}[htbp]
\centering
\includegraphics [width=0.4\textwidth] {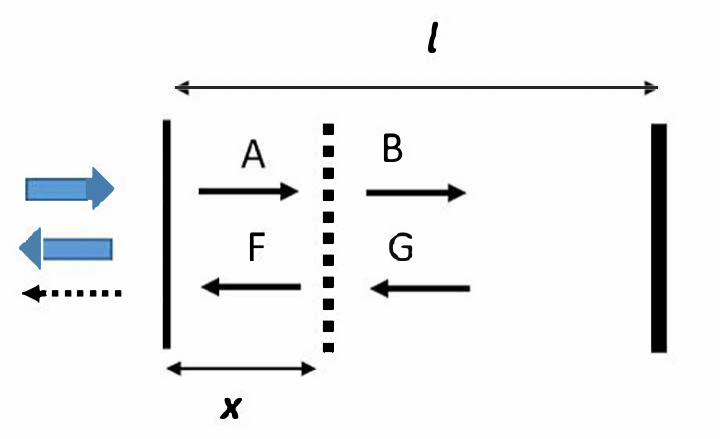}
\caption{A one-sided cavity with a mechanical semitransparent membrane set inside it. The dashed line
represents the membrane, the thin solid line represents the semitransparent input mirror, the thick solid line represents the perfectly reflecting back-stop mirror, the thick arrows represent the pump light, and the dashed arrow represents the detected light.
 The waves inside the cavity are shown with arrows which are marked with their complex amplitudes.
\label{FF1}}
\end{figure}

An optical cavity with a mechanically compliant membrane placed inside it is a widely used optomechanical setup. Its typical realization is the so-called "membrane-in-the-middle" cavity (MIM), which consists of  a Fabry-Perot optical cavity where the membrane is placed  close to its middle~\cite{jayich_dispersive_2008, miao_standard_2009, yanay_quantum_2016, thompson_strong_2008, wilson_cavity_2009, purdy_strong_2013, mason_continuous_2019, kampel_improving_2017, higginbotham_harnessing_2018}.
In terms of Fig.~\ref{FF1}, for such a configuration, the distance $x$ is set close to $l/2$.
Recently, it was identified~\cite{dumont_flexure-tuned_2019} that, if the membrane is highly reflecting and properly positioned close to the input mirror, the optomechanical coupling constant of the system  can be appreciably increased compared to the MIM configuration.
Such a configuration was called  "membrane-at-the-edge"(MATE). In Ref.~\cite{dumont_flexure-tuned_2019} , it was shown that, for MATE, there exists a regime where the optomechanical coupling constant increases inversely proportional to the membrane transmission. 
Since, currently, mechanical membranes with an extremely low transmission are available
\cite{chen2017high,enzian2023phononically,zhou_2023_cavity}, an efficient use of highly reflecting membranes in optomechanical cavity setups makes an issue of an appreciable interest.

In this paper, we theoretically address the configuration where a highly reflecting  membrane is positioned close to the backstop mirror, which we call  "membrane-at-the-back" (MAK).
We will compare the optimized optomechanical performance of the three aforementioned configurations using the optomechanical cooperativity, as well as the efficiency-weighted cooperativity as figures of merit.
Only the dispersive optomechanical coupling, as typically dominating the systems, is taken into account.

In contrast to the previous  theoretical considerations of an optical cavity with a membrane inside \cite{jayich_dispersive_2008, miao_standard_2009,dumont_flexure-tuned_2019}, we have incorporated the parasitic scattering from the membrane in our consideration while neglecting the parasitic scattering from the coupling mirror, which we assume to be much weaker than that from the membrane.
We have demonstrated that, in the case of a  highly reflecting membrane, even a small amount of parasitic scattering from the membrane may have an essential impact on the cooperativity of the system.

In the case of a highly reflecting  membrane, the spectrum of the cavity is close to the  superposition of those of the two subcavities, being strongly affected near the crossing points of these spectra by the effect of avoided crossing.
At the avoided-crossing  points,  the dispersive optomechanical coupling  vanishes.
Evidently, halfway between these  points, the suppressing effect of the avoided crossing on the dispersive  coupling is minimal.
Being interested in the best performance of the system, the consideration in the paper will be focused  on these "halfway" points.

We will show that, for highly reflecting membranes, in terms of the cooperativity, MAK is always appreciably  advantageous compared  to  MIM while, depending on the parameters of the  system, MAK is either appreciably  advantageous compared  to  MATE or yields a performance which is very close to that of MAK.

 We will also compare the MAK scheme against the simple Fabry-Perot cavity (FP), with the highly reflecting membrane in the role of the coupling mirror.
 We will show that, depending on the parameters of the systems, MAK can be more or less advantageous than the aforementioned  FP  of the length equal to the membrane/mirror separation in MAK.
\section{System model}
To be specific, we set the following scattering matrices:
\begin{equation}
\label{mirror}
 \left(
  \begin{array}{cc}
 it &  -r \\
-r  & it  \\
  \end{array}
\right)\qquad
\end{equation}
for the input mirror,
\begin{equation}
\label{mirrorB}
 \left(
  \begin{array}{cc}
 0 &  -1 \\
-1  & 0  \\
  \end{array}
\right)\qquad
\end{equation}
for the backstop mirror, and
\begin{equation}
\label{membrane}
 \left(
  \begin{array}{cc}
 t_me^{i\varphi_t} &  r_me^{i\varphi_r} \\
r_m e^{i\varphi_r} & t_me^{i\varphi_t}  \\
  \end{array}
\right)\qquad
\end{equation}
for the  membrane. In matrices (\ref{mirror}), (\ref{mirrorB}) and (\ref{membrane}), the amplitude transmission coefficients are on the diagonals. 
Here all parameters, including the phases $\varphi_t$ and $\varphi_r$, are set as real and positive while 
\begin{equation}
\label{parameters}
 t^2+r^2 =1\mathrm\qquad {\mathrm{and}}\qquad  e^{2i(\varphi_r-\varphi_t)}=-1.
\end{equation}
The finesse of the cavity is assumed to be high:
\begin{equation}
\label{t}
 t\ll 1
\end{equation}
Though the treatment of cavities with a membrane inside is typically done neglecting parasitic scatting of the membrane, i.e. assuming  $t_m^2+r_m^2 =1$, we  will incorporate it into the consideration since as will be seen below such a scattering may strongly affect the behavior of the system addressed.
Specifically we set
\begin{equation}
\label{parameters1}
 t_m^2+r_m^2 =1- t_s^2
\end{equation}
where $t_s^2$ is the power scattering coefficient associated with  the aforementioned scattering.

Throughout the paper we will restrict ourselves to the case of high reflecting membrane,  i.e.
\begin{equation}
\label{ineq}
 t_m^2\ll 1.
\end{equation}

As it is seen from Eqs.(\ref{mirrorB}) and (\ref{parameters}) we neglect the parasitic scattering against the coupling mirror and a finite transmission of the backstop mirror.
This is justified under reasonable assumptions that the above scattering is much weaker than the  coupling mirror transmission while  the power transmission of the backstop is much smaller than $t_s^2$.
\section{Basic features of an optical cavity with a highly reflecting membrane inside}
\label{basic}
In this section, we address some basic features of an optical cavity with a highly  reflecting membrane inside.
We will do it neglecting the dissipation effects 
in the system.

\begin{figure}[htbp]
\centering
\includegraphics[width=0.6\textwidth]{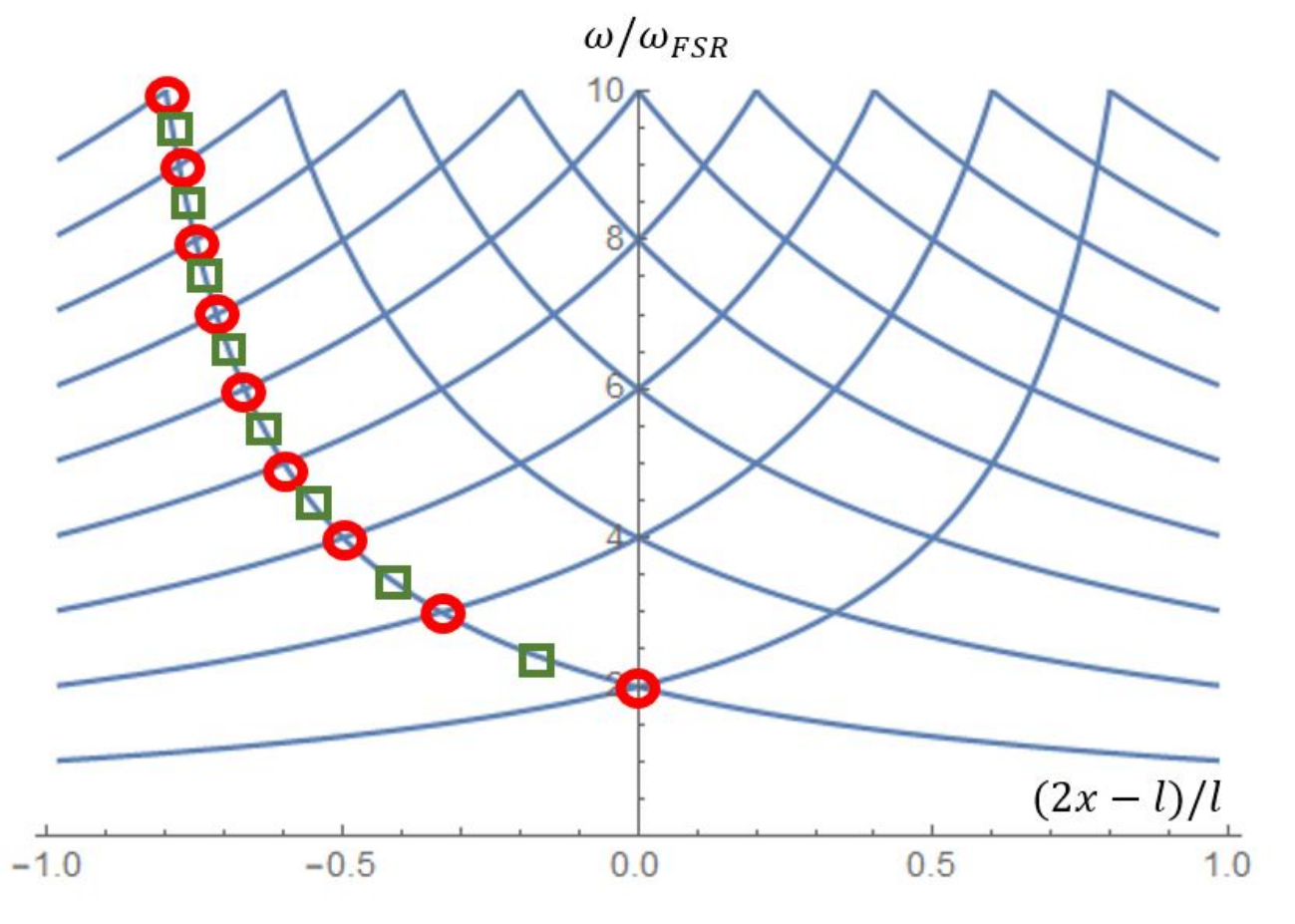}
\caption{
The spectrum of an optical cavity with a perfectly reflecting membrane inside ($r_m=1$)  as a function of the membrane position.
The phase shift at reflection from the membrane is set as $\varphi_r =\pi$.
$\omega_{\mathrm{FSR}}=\pi c/l$ is the free spectral range of the empty cavity where $l$ is the cavity length and $c$ is the speed of light.  
For a selected  $\omega_\mathrm{L}$ mode, the places where it crosses a number of the $\omega_\mathrm{R}$ modes are
marked with the circles.  The r-points, where the selected $\omega_\mathrm{L}$ is on resonance while the mode corresponding  $\omega_\mathrm{R}$ is on anti-resonance, are marked with the squares.
\label{FF2}}
\end{figure}

\begin{figure}[htbp]
\centering
\includegraphics [width=0.5\textwidth]{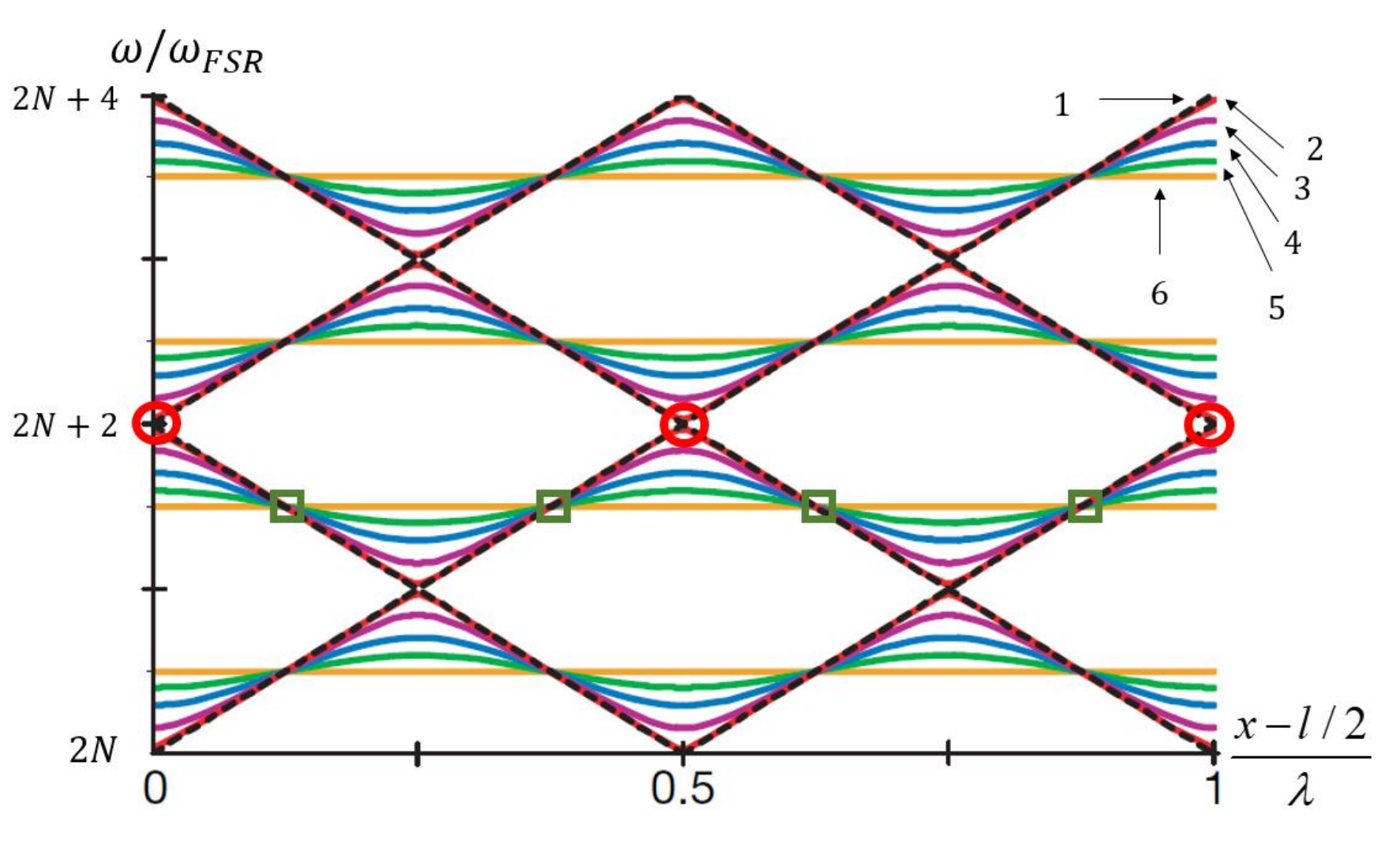}
\caption{A part of the spectrum of an optical cavity with a semitransparent  membrane inside for high order cavity resonances ($N\gg 1$ is an integer) plotted for different values of its amplitude transmission coefficients $t_m$:
1 (1), 0.95 (2), 0.8 (3), 0.48 (4), 0.1 (5), and 0 (6).
$\omega_{\mathrm{FSR}}$ is the free spectral range of the empty cavity.
The phase shift at reflection from the membrane is set as $\varphi_r =\pi$.
Three crossing points are marked with circles.
Four "r" points are marked with squares.
Figure is based on the results from  Ref. \cite{thompson_strong_2008}.
\label{FF3}}
\end{figure}

Let us start from the case of a perfectly reflecting membrane ($t_m=0$).
In this case, the resonance spectrum of the system for any position of the membrane is just the sum of spectra of its right and left sub-cavities as shown in  Fig.~\ref{FF2}.
The resonance frequencies of modes of the left part $\omega_\mathrm{L}$ decrease with increasing $x$ while those of modes of the right part $\omega_\mathrm{R}$ increase.
Hereafter we will term such modes  $\omega_\mathrm{L}$ modes and $\omega_\mathrm{R}$ modes, respectively.
For small but finite $t_m$, the spectrum shown in Fig.~\ref{FF2} provides a good approximation for the spectrum except the vicinity of the crossing points where an avoided crossing takes place, resulting in the formation of a small gap.
Away from avoided crossing points, we can still keep unambiguous nomenclature, $\omega_\mathrm{L}$ mode and $\omega_\mathrm{R}$ mode, to classify the modes.
The modification of the spectrum caused by a variation of $t_m$ is illustrated in Fig.~\ref{FF3}.
Here the effect of avoided crossings at the crossing points is seen.
Three crossing points are marked with circles.
An important feature of the avoided crossing is that this effect is the stronger, the stronger the displacement of the membrane from the middle of the cavity is.
The strength of this effect can be quantified by the frequency gap $\delta\omega$, which appears at the crossing point.
Using the following well-known equation
\begin{equation}
\label{fr}
\cos(kl+\varphi_r) = -r_m\cos(2kx-kl),
\end{equation}
for the resonant wave vector $k$ of an optical cavity with a semitransparent mirror inside (see e.g.~\cite{dumont_flexure-tuned_2019} ), which is written neglecting the energy decay in the cavity, in the limit $ t_m\ll 1$, one finds (see Appendix~\ref{Gap} )
\begin{equation}
\label{gap0}
\delta \omega\propto  \frac{ct_m}{\sqrt{x_0(l-x_0)}},
\end{equation}
where $x_0$ is the distance from a crossing point to the left mirror of the cavity and $c$ is the speed of light.
Equation (\ref{gap0}) justifies the above statement.

Solution to Eq.(\ref{fr}) also yields (c.f. Eqs.~(\ref{set}))
\begin{equation}
\label{om0}
\omega_0= \omega_{\mathrm{FSR}} \left(N-\frac{\varphi_r}{\pi} \right)
\end{equation}
for the frequency of the crossing points $\omega_0$, where $\omega_{\mathrm{FSR}}=\pi c/l$ is the free spectral range of the empty cavity, $l$ is the cavity length,  and $N$ is an integer.

Another remarkable feature of the system, which is seen from Fig.~ \ref{FF3}, is that the positions of the points where the dispersion curves of the decoupled modes (the straight lines in the figure) cross the dispersion curves of the system (at finite membrane transmission $t_m$) are not sensitive to the value of $t_m$ (see Appendix \ref{rpoints}).
In Fig.~\ref{FF3}, four of such points are marked with squares.
With a variation of $t_m$, the $\omega(x)$ curves ``locally rotate" about such points.
We term these points r-points.

Solution to Eq.(\ref{fr}) yields (c.f. Eqs.~(\ref{Lresonance}),  (\ref{Rresonance}), and  (\ref{mu})) 
\begin{equation}
\label{omr}
\omega_r= \omega_{\mathrm{FSR}} \left(\frac{ 1}{2}+N-\frac{\varphi_r}{\pi} \right)
\end{equation}
for the frequency of the r-points.

Figure \ref{FF2} illustrates the relative position of the points of avoided crossing and r-points on an $\omega_\mathrm{L}$ branch of the spectrum.
It is seen that, in  frequency, these are separated by $\omega_{\textrm{FSR}}/2$ as it follows from equations~(\ref{om0}) and (\ref{omr}).
In other words, each r-point lies in the middle between two neighboring points of avoided crossing.
The avoided crossing obviously results in a reduction of the slopes of the $\omega(x)$ curves and, as a result,  in a reduction of the optomechanical coupling.
It is clear that the middle position of the r-points between the neighboring points of avoided crossing makes the r-points favorable in terms of such a coupling.

For this reason, in the further discussion, we will focus on these points.
Such a discussion will be based on the following properties of the system at the r-points.
The derivation of these properties is given in Appendix \ref{rpoints}. Here we just list them.

At the r-points one of the sub-cavities is on resonance while the other is on antiresonance.
Mathematically, in terms of the resonance wavevectors $k$, this means that, for    $\omega_\mathrm{L}$ modes:
\begin{equation}
\label{resL}
e^{-2ik_0(l-x) -i\varphi_r}=1,\qquad  e^{-2ik_0x -i\varphi_r}=-1.
\end{equation}
while for $\omega_\mathrm{L}$ modes: 
\begin{equation}
\label{resR}
e^{-2ik_0(l-x) -i\varphi_r}=-1,\qquad  e^{-2ik_0x -i\varphi_r}=1.
\end{equation}

For the r-points and the case of highly reflecting membranes, the ratio of the field intensity in one part of the cavity to that in the other part is fully controlled by the  power (intensity) transmission of the membrane $T_m=t_m^2$.
Specifically, for the $\omega_\mathrm{L}$ mode, 
\begin{equation}
\label{ratio0}
\frac{|B|^2}{|A|^2}=\frac{T_m}{4},
\end{equation}
where  $|A|^2$ and $|B|^2$ are the  intensities in the x-long and l-x-long parts of the cavity, respectively. For $\omega_\mathrm{R}$ mode, we have
\begin{equation}
\label{ratio1}
\frac{|B|^2}{|A|^2}=\frac{4}{T_m}.
\end{equation}.

As seen from Fig.~\ref{FF3}, once the membrane is close to the middle of the cavity, for small  $t_m$, the slope of the $\omega(x)$  curves at the r-points is hardly affected by the coupling between the sub-cavities.
However, as was shown above, the effect of such coupling increases when the membrane approaches one of the cavity mirrors.
Thus, for small distances between the membrane and the mirror, the impact of this coupling  on the aforementioned slope may become appreciable.
Let us introduce the critical membrane-mirror separation, we term it $x_\textrm{int}$, at which this happens.
We mean that, at $x_\textrm{int}\ll x\ll l$  the effect on the inter-cavities coupling of the spectrum near the r-points is negligible while, at $ x \ll x_\textrm{int}\ll l$, it is very strong.
These two regimes can also be viewed in terms of the mode energy.
At $x_\textrm{int}\ll x\ll l$, the sub-cavities are virtually decoupled such that the energy of the $\omega_\mathrm{L}$ mode, as that originating from the left smaller sub-cavity, should be  mainly stored in this sub-cavity.
On the other hand, for $ x \ll x_\textrm{int}\ll l$,
since now the inter-cavities coupling is very strong, the main fraction of energy of the $\omega_\mathrm{L}$ mode should be stored in the right sub-cavity.
In view of the above-said we determine $x_\textrm{int}$ from the condition that, at $x=x_\textrm{int}$, the energy of the $\omega_\mathrm{L}$ mode is equally distributed between the two sub-cavities.
This condition readily yields 
(see Appendix~\ref{rpoints}) that
\begin{equation}
\label{half0}
x_{\textrm{int}}=l\frac{T_m}{4}.
\end{equation}
where $T_m=t_m^2$ is the  power (intensity) transmission of the membrane.

Evidently, for the $\omega_\mathrm{L}$ mode, $x_\textrm{int}$ also represents the corresponding critical separation between the membrane and the backstop mirror. 
\section{Dispersive coupling for an optical cavity with a highly reflecting membrane inside}
\label{coupling}
In this section, we consider the dispersive optomechanical coupling at the r-points.
Since we are interested in the case of weak dissipation effects, in this section, we evaluate the spectrum of the system neglecting these effects.  

We define the dispersive coupling constant of the system as follows
\begin{equation}
\label{g0}
g_0= -\frac{d\omega_c}{dx}x_{\textrm{zpf}}
\end{equation}
where $\omega_c$ is the resonance frequency of the system and $x_{\textrm{zpf}}$ is the amplitude of zero-point fluctuations.
Let us first address $g_0$ of the $\omega_\mathrm{L}$ mode.
According to Sec.~\ref{basic}, at $l- x_\textrm{int} \gg x\gg x_\textrm{int}$, the dispersive coupling for the decoupled $\omega_\mathrm{L}$ mode, i.e. at $t_m=0$, provides a good approximation for that at finite $t_m$, implying
\begin{equation}
\label{out}
g_{0}=\frac{\omega_c}{x}x_{\textrm{zpf}}.
\end{equation}
Here, the $1/x$ increase of $g_0$  at $x\rightarrow 0$ results from the confinement of the mode energy in a decreasing volume.
At $x$ approaching  $x_\textrm{int}$, this trend saturates such that one can reasonably suppose that at $x \ll x_\textrm{int}$
\begin{equation}
\label{in}
g_{0}=\frac{\omega_c}{ x_\textrm{int}}x_{\textrm{zpf}},\qquad  x_{\textrm{int}}=l\frac{T_m}{4}.
\end{equation}
The result given by (\ref{in}) is consistent with the enhanced value of $g_0$, which was identified for the system in reference~\cite{dumont_flexure-tuned_2019}.

Such a heuristic result is readily supported by direct calculations.
We rewrite  the resonance equation  (\ref{fr}) in the following equivalent form (see Appendix \ref{fullset})
\begin{equation}
\label{resN}
(r_m +e^{-2ik(l-x) -i\varphi_r})(r_m +e^{-2ikx -i\varphi_r})+1- r_m^2 = 0
\end{equation}
and calculate $\frac{dk}{dx}$ at $k$ satisfying Eq.(\ref{resL}). 
Next, using (\ref{g0}) and taking into account that $T_m\ll1$ we find  (see Appendix \ref{fullset})
\begin{equation}
\label{L}
g_{0}= \frac{\omega_c}{x+x_{\textrm{int}}}x_{\textrm{zpf}},
\end{equation}
justifying (\ref{in}).

The above heuristic argument holds for the $\omega_\mathrm{R}$ mode while, the direct calculation involving   (\ref{resN})  and (\ref{resR})  yields (see Appendix \ref{fullset})
\begin{equation}
\label{R}
g_{0}= -\frac{\omega_c}{l-x+x_{\textrm{int}}}x_{\textrm{zpf}}.
\end{equation}

Next, basing on equations (\ref{L}) and (\ref{R}), in Table~\ref{Tab1}, we summarize the values of $|g_0|$ for three regimes,  where $x\ll x_{\mathrm{int}}$, $|x-l/2|\ll l$, and $l-x\ll x_{\mathrm{int}}$, which we label MATE, MIM, and MAK, for the membrane-at-the-edge, membrane-in-the-middle, and membrane-at-the back systems, respectively.
In this Table,  the  values of $|g_0|$ given for MATE with the $\omega_\mathrm{L}$ mode and for MAK with the $\omega_\mathrm{R}$ mode are the same, being the upper limit  for $|g_0|$ in this system.   
The values  for MATE with the $\omega_\mathrm{L}$ mode and for MAK with the $\omega_\mathrm{R}$ mode given in this table hold to within a factor of $1/2$ also for results for the membrane/mirror separation equal to  $x_{\mathrm{int}}$.
\begin{table}[htbp]
\centering
\begin{tabular}{|c|c|c|c|}
\hline
 & MATE & MIM & MAK  \\
\hline
$\omega_\mathrm{L}$ & $2/T_\textrm{m}$&  $1$  & $1/2$ \\
\hline
$\omega_\mathrm{R}$ & $1/2$ & $1$ & $2/T_\textrm{m}$ \\
\hline
\end{tabular}
\caption{\label{Tab1}The absolute values of the dispersive optomechanical coupling constants normalized to the MIM coupling constant $g_0/(2 \omega_c x_\textrm{zpf}/l  )$ for the $\omega_\mathrm{L}$ and $\omega_\mathrm{R}$ modes at r-points for  three configurations:
$x\ll x_{\mathrm{int}}$, $|x-l/2|\ll l$, and $l-x\ll x_{\mathrm{int}}$, which are labeled as MATE, MIM, and MAK, respectively.
 $T_m$ is the power transmission coefficient of the membrane.
 The values  for MATE with the $\omega_\mathrm{L}$ mode and for MAK with the $\omega_\mathrm{R}$ mode given in this table to within a factor of $1/2$ hold for the results for the membrane/mirror separation equal to  $x_{\mathrm{int}}$. 
}
\end{table}
Table~\ref{Tab1} suggests that, w.r.t. dispersive coupling, MATE and MAK yield a performance which, for highly reflecting membranes, is substantially superior to that of MIM.


\section{Cavity linewidth and cooperativity for an optical cavity with a highly reflecting membrane inside }

The dispersive coupling constant $g_0$ evaluated above does not represent, in general, a reliable optomechanical figure of merit.
An appropriate figure of merit for mechanical sensing \cite{wilson_measurement-based_2015} and optomechanical squeezing
\cite{tagantsev_dissipative_2018} is the so-called single-photon cooperativity,
which reads
\begin{equation}
\label{C}
C = \frac{4g_0^2}{\kappa \gamma_m},
\end{equation}
where $\kappa$ is the optical decay rate and $\gamma_m$ is the mechanical decay rate.
$C$ is a fully appropriate parameter in the case where the linewidth of the system is controlled only by the energy leakage though the coupling mirror.
However, once the parasitic scattering against the membrane is involved, after stating that the linewidth can be decomposed into a contribution from external coupling and parasitic loss
\begin{equation}
\label{kappa}
\kappa=\kappa_{\mathrm{ext}}+\kappa_{s} \ ,
\end{equation}
we can define the efficiency-weighted cooperativity
\begin{equation}
\label{Cw}
C_\eta = \eta C = \frac{4g_0^2\kappa_{\mathrm{ext}}}{(\kappa_{\mathrm{ext}}+\kappa_{s})^2\gamma_m}
\end{equation}
 where we introduced the coupling efficiency $\eta$ as
\begin{equation}
\label{eta}
\eta = \frac{\kappa_\text{ext}}{\kappa} .
\end{equation}
One readily checks that, for optical sensing of the mechanical subsystem, it is the efficiency-weighted cooperativity that plays a role of the figure of merit.

\subsection{Simple estimates and qualitative arguments }
Prior to a more detailed analysis, we compare the cooperativity performance of MATE, MIM, and MAK in the standard framework  \cite{jayich_dispersive_2008}~\cite{dumont_flexure-tuned_2019} with the parasitic scattering of the membrane  neglected, i.e. we set $t_s=0$, c.f. Eq.( \ref{parameters1}).
We do it for the r-points and for the regime where MATE and MIM yield the maximum $g_0$ , i.e. for the membrane separations from the adjacent mirror, which are much smaller than $x_{\textrm{int}}=l\frac{T_m}{4}$.

\begin{figure}[htbp]
\includegraphics [width=0.7\columnwidth,clip=true, trim=0mm 0mm 0mm 0mm] {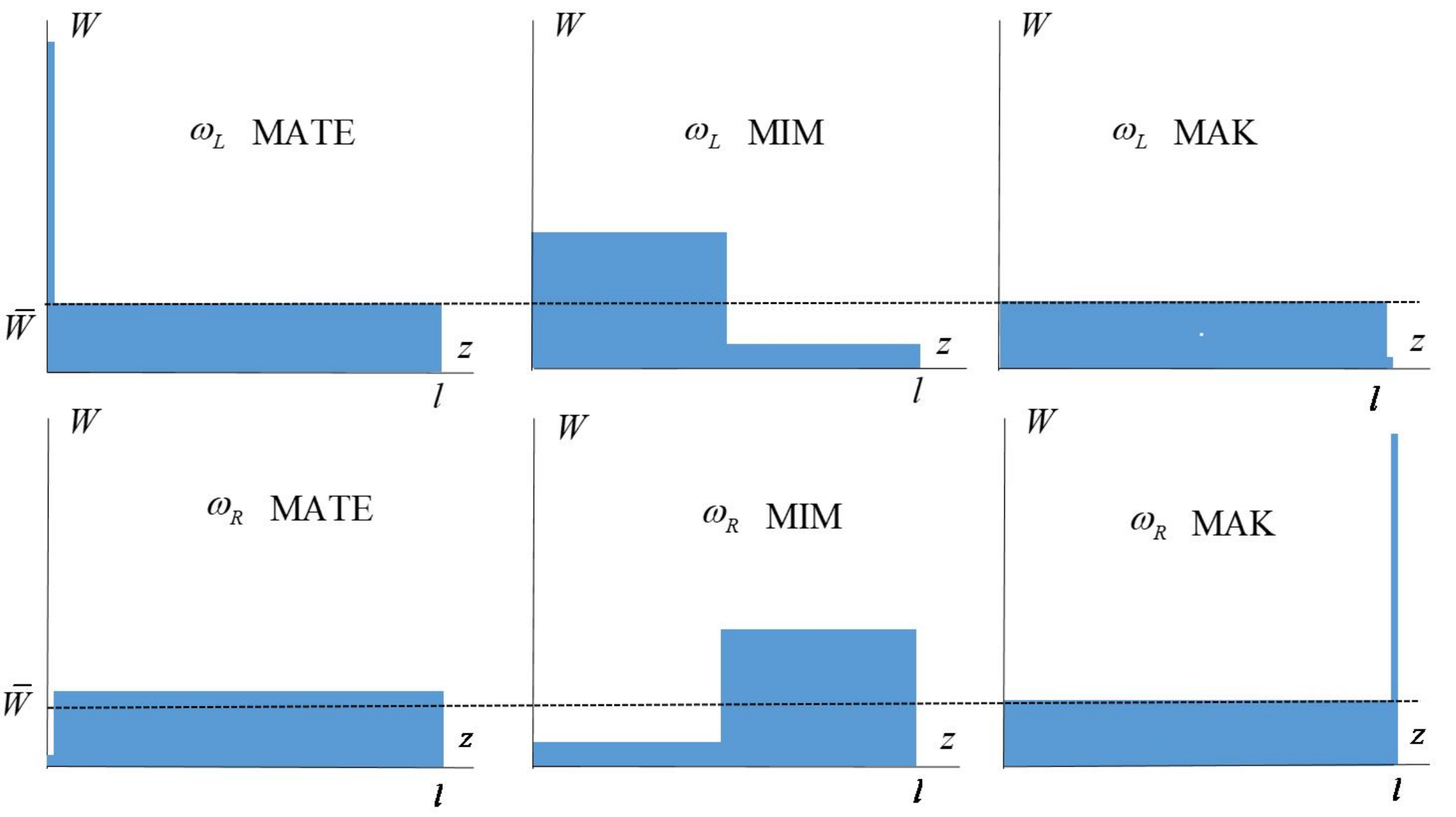}
\caption{The distribution of the field intensity in the MATE, MIM, and MAK configurations at r-points for the $\omega_\mathrm{L}$ and $\omega_\mathrm{R}$ modes.
$\bar{W}$ denotes the field intensity averaged over the cavity.
 For MATE and  MAK, the regime where the membrane separation from the adjacent  mirror is much smaller than $x_{\textrm{int}}=l\frac{T_m}{4}$ is considered.
Figure is not to scale.
\label{FF5}}
\end{figure}

The optical decay rate of MATE, MIM, and MAK can be readily found from  the field distributions in these systems shown in Fig.\ref{FF5} and the standard definition (see e.g.~\cite{dumont_flexure-tuned_2019}) of the decay rate:
\begin{equation}
\label{kappa1}
\kappa = \frac{\mathrm{dissipated\,\,power}}{\mathrm{stored\,\, energy} } = \frac{ct^2W(0)}{2l\bar{W}},
\end{equation}
where $W(0)$ and $\bar{W}$ are the optical field intensity at the input mirror and the optical field intensity averaged over the cavity.
Since the case of weak dissipative effects is addressed, Eqs. (\ref{ratio0}) and  (\ref{ratio1}), which were obtained neglecting the dissipation, can be used for the evaluation of $W(0)$ and $\bar{W}$. 
Figure \ref{FF5} schematically illustrates the  relations between the field intensities given by  Eqs. (\ref{ratio0}) and  (\ref{ratio1}).
For MIM, using (\ref{ratio0}) and (\ref{ratio1}), we can write:
\begin{equation}
\label{MIMdr}
  \begin{array}{cc}
W(0)\approx2\bar{W}\qquad \mathrm{for}\qquad \omega_L,\\
W(0) \approx \frac{T_m}{2}\bar{W}\qquad \mathrm{for}\qquad \omega_R. \\
  \end{array}
\end{equation}
As for MATE and MAK, taking into account that, in the addressed regime where the membrane separation from the adjacent mirror is much smaller than $x_{\textrm{int}}$, the energy of the mode is mainly stored in the larger subcavities, one finds that for MATE
\begin{equation}
\label{MATEdr}
  \begin{array}{cc}
W(0) \approx\frac{4}{T_m}\bar{W}\qquad \mathrm{for}\qquad \omega_L,\\
W(0) \approx\frac{T_m}{4}\bar{W}\qquad \mathrm{for}\qquad \omega_R. \\
\end{array}
\end{equation}
while, in MAK, for both modes
\begin{equation}
\label{MAKdr}
W(0) \approx\bar{W}.
\end{equation}
Next, using the above results, those from Table \ref{Tab1}, and  Eqs. (\ref{kappa1}) and  (\ref{C}) we arrive at the results for the cavity linewidth and cooperativity $C$ , which are summarized in Tables \ref{Tab2} and  \ref{Tab3}.
\begin{table}[htbp]
\centering
\begin{tabular}{|c|c|c|c|}
\hline
 & MATE & MIM & MAK  \\
\hline
$\omega_\mathrm{L}$ & $2/T_m$&  1  & $1/2$ \\
\hline
$\omega_\mathrm{R}$ & $T_m/8$ & $T_m/4$ & $1/2$ \\
\hline
\end{tabular}
\caption{\label{Tab2}The normalized cavity linewidth $\kappa/\kappa_0$, $\kappa_0= cT/l$, for the $\omega_\mathrm{L}$ and $\omega_\mathrm{R}$ modes at three configurations:
 $x\ll x_{\mathrm{int}}$, $|x-l/2|\ll l$, and $l-x\ll x_{\mathrm{int}}$, which are labeled as MATE, MIM, and MAK, respectively.
  $T $ and $T_m $ are the power transmission coefficients of the coupling mirror and membrane, respectively.
 The values  for MATE with the $\omega_\mathrm{L}$ mode and for MAK with the $\omega_\mathrm{R}$ mode given in this table to within a factor of $1/2$ hold for the results for the membrane/mirror separation equal to  $x_{\mathrm{int}}$.
 The results are valid for the  r-points of the spectrum.
 }
\end{table}
\begin{table}[htbp]
\centering
\begin{tabular}{|c|c|c|c|}
\hline
 & MATE & MIM & MAK  \\
\hline
$\omega_\mathrm{L}$ & $2/T_m$&  1  & $1/2$ \\
\hline
$\omega_\mathrm{R}$ & $2/T_m$ & $4/T_m$ & $8/T_m^2$ \\
\hline
\end{tabular}
\caption{\label{Tab3} The normalized cooperativity $ C/C_0$, $C_0= 16(\omega_c x_{\textrm{zpf}})^2/(T cl\gamma_m)$, for the $\omega_\mathrm{L}$ and $\omega_\mathrm{R}$ modes at three configurations:
 $x\ll x_{\mathrm{int}}$, $|x-l/2|\ll l$, and $l-x\ll x_{\mathrm{int}}$, which are labeled as MATE, MIM, and MAK, respectively.
 $T $ and $T_m $ are the power transmission coefficients of the coupling mirror and membrane, respectively.
 The values  for MATE with the $\omega_\mathrm{L}$ mode and for MAK with the $\omega_\mathrm{R}$ mode given in this table to within a factor of $1/2$ hold the for results for the membrane/mirror separation equal to  $x_{\mathrm{int}}$.
 The results are valid for the  r-points of the spectrum.
 }
\end{table}
From Table \ref{Tab3} one concludes that for an ideal membrane, i.e. a membrane exhibiting no parasitic scattering, w.r.t. the cooperativity, despite higher $g_0$  MATE is not advantageous compared to MIM while  MAK is strongly advantageous (at least $2/T_m$ times) compared to other  systems.

Another conclusion can be drawn from scrutinizing Fig.\ref{FF5}.
 It is seen that, for the $\omega_\mathrm{R}$ mode in MIM and MAK, which, according to Table \ref{Tab3},  are the two  most advantageous cases, the reflecting membrane is in contact with a field, which is  much larger than that at the coupling mirror.
 This suggests that even very small parasitic scattering may be detrimental for the linewidth and cooperativity for these systems.
 In the next Subsection, we will address this matter in detail.
\subsection{Linewidth and cooperativity in the presence of parasitic scattering }
In order to calculate the linewidth of the system, one can generalize Eq. (\ref{fr}), which is  written for the real resonance  wave vector, to the following resonance equation for the complex wave vector $k$ (see Appendix \ref{fullset})
\begin{equation}
\label{resDISS}
(r_m +e^{-2ik(l-x) -i\varphi_r})(r_m +r^{-1}e^{-2ikx -i\varphi_r})+1- r_m^2 = T_s
\end{equation}
where $T_s = t_s^2$ is the power scattering coefficient associated with  the parasitic scattering (see Eq.(\ref{parameters1})).
Next, one finds the linewidth $\kappa $ as follows
\begin{equation}
\label{LW}
\kappa =-2c\mathrm{Im}[k].
\end{equation}

Being interested in the $r$-points of the spectrum in the regime of small damping, we are looking  for a solution to (\ref{resDISS})  written in the form: $k=k_0 +\delta k$ where $k_0$ is real solution to Eq.~(\ref{fr}) at these points.
Since the situation of a weak dissipation is addressed $|\delta k|\ll |k_0|$ such that $\delta k$ can be calculated by linearizing Eq. (\ref{resDISS}).
This way and using (\ref{LW}) (see Appendix \ref{fullset}) for the  $\omega_\mathrm{R}$  mode  one finds
\begin{equation}
\label{dampingRe}
\kappa_{\mathrm{ext}} = \frac{c}{2}\frac{TT_m/4}{l-x +x_{\textrm{int}}},
\end{equation}
\begin{equation}
\label{dampingRs}
\kappa_s = \frac{c}{2}\frac{T_s}{l-x +x_{\textrm{int}}} ,
\end{equation}
implying for the efficiency:
\begin{equation}
\label{etaR}
\eta= \frac{TT_m/4}{T_s+TT_m/4} .
\end{equation}
While, for the $\omega_\mathrm{L}$ mode:
\begin{equation}
\label{dampingLe}
\kappa_{\mathrm{ext}}=  \frac{c}{2}\frac{T}{x +x_{\textrm{int}}},
\end{equation}
\begin{equation}
\label{dampingLs}
\kappa_s =  \frac{c}{2}\frac{T_s}{x +x_{\textrm{int}}},
\end{equation}
implying for the efficiency:
\begin{equation}
\label{etaL}
\eta= \frac{T}{T_s+T} .
\end{equation}
Here, $T = t^2$ is the power transmission coefficient of the coupling mirror.

One readily notes that the cooperativity calculated neglecting the parasitic scattering against the membrane should be  multiplied by the efficiency to yield  the cooperativity calculated taking it into account.
Thus using Table~\ref{Tab3} and Eqs.~(\ref{etaR}) and (\ref{etaL}) we arrive at the results listed in Table~\ref{Tab4}.
\begin{table}[htbp]
\centering
\begin{tabular}{|c|c|c|c|}
\hline
 & MATE & MIM & MAK  \\
\hline
$\omega_\mathrm{L}$ & $\frac{2T}{T_m(T_s+T)}$& $\frac{T}{T_s+T}$  & $\frac{T/2}{T_s+T}$ \\
\hline
$\omega_\mathrm{R}$ & $\frac{T/2}{T_s+TT_m/4}$  & $\frac{T}{T_s+TT_m/4}$ & $\frac{2T}{T_m(T_s+TT_m/4)}$ \\
\hline
\end{tabular}
\caption{\label{Tab4} Cooperativity calculated taking into account the parasitic of the membrane.  The normalized cooperativity $ C/C_0$, $C_0= 16(\omega_c x_{\textrm{zpf}})^2/(T cl\gamma_m)$, for the $\omega_\mathrm{L}$ and $\omega_\mathrm{R}$ modes at three configurations:
 $x\ll x_{\mathrm{int}}$, $|x-l/2|\ll l$, and $l-x\ll x_{\mathrm{int}}$, which are labeled as MATE, MIM, and MAK, respectively.
  Here $T_s = t_s^2$, $T = t^2$, and  $T_m = t_m^2$,   are the power scattering coefficient associated with  the parasitic scattering and  the power transmission coefficients of the coupling mirror and membrane, respectively.
 The values  for MATE with the $\omega_\mathrm{L}$ mode and for MAK with the $\omega_\mathrm{R}$ mode given in this table to within a factor of $1/2$ hold for the results for the membrane/mirror separation equal to  $x_{\mathrm{int}}$.
 The results are valid for the  r-points of the spectrum.}
\end{table}

The results presented in in Table~\ref{Tab4} can be summarized as  follows.
We will  discuss the $\omega_\mathrm{R}$-regime of MAK, the $\omega_\mathrm{R}$-regime of MIM, and the $\omega_\mathrm{L}$-regime of MATE.
The ``standing" of those regimes  depends of the position of  $T_s$ with respect to $TT_m/4$ and $T$.
At $T_s\ll TT_m/4$, we are back to the dissipation free regime and MAK is  advantageous compared to other regimes by a factor of about $1/T_m$.
At $ TT_m/4\ll T_s\ll T$,  MAK is  advantageous compared to MIM  by a factor of about $1/T_m$ being advantageous compared to MATE  by a factor of about $T/T_s$.
And finally, at $T_s\gg T$,  MAK and MATE yields practically the same cooperativity  being advantageous compared to MIM  by a factor of about $1/T_m$.
All in all, it is seen that, once the parasitic scattering against the membrane is taken into account, in terms of the cooperativity, MAK is the best though, at $T_s\gg T$, the performance of MATE is very close to that of MAK.

It is instructive to give general expressions for the cooperativity of  the $\omega_\mathrm{R}$  and $\omega_\mathrm{L}$ modes. 
Using Eqs.~(\ref{L}), (\ref{R}), (\ref{C}), (\ref{kappa}), (\ref{dampingRe}), (\ref{dampingRs}), (\ref{dampingLe}), and (\ref{dampingLs}) we find
\begin{equation}
\label{CL}
C=  \frac{8}{c\gamma_m}\frac{(\omega_c x_{\textrm{zpf}})^2}{x +x_{\textrm{int}}}\frac{1}{T_s +T}
\end{equation}
for the $\omega_\mathrm{L}$ mode and
\begin{equation}
\label{CR}
C=  \frac{8}{c\gamma_m}\frac{(\omega_c x_{\textrm{zpf}})^2}{l-x +x_{\textrm{int}}}\frac{1}{T_s +TT_m/4}
\end{equation}
for the $\omega_\mathrm{R}$ mode.

It is clear from these expressions, that, passing from the optimal regime for MATE with the $\omega_\mathrm{L}$
mode and for MAK with the $\omega_\mathrm{R}$ where the mirror/membrane separation $\delta$ much smaller than $x_{\textrm{int}}$ to the regime with $\delta=x_{\textrm{int}}$, one looses in the cooperativity only a factor of 2.
\subsection{Efficiency-weighted   cooperativity}
Multilying Eqs.~(\ref{CL}) and (\ref{CR}) with the efficiency given by Eqs.~(\ref{etaL}) and (\ref{etaR}) we find
\begin{equation}
\label{CLe}
C_\eta= \frac{8}{c\gamma_m}\frac{(\omega_c x_{\textrm{zpf}})^2}{x +x_{\textrm{int}}}\frac{T}{(T_s +T)^2}
\end{equation}
for the efficiency-weighted cooperativity of the $\omega_\mathrm{L}$ mode  and
\begin{equation}
\label{CRe}
C_\eta=  \frac{8}{c\gamma_m}\frac{(\omega_c x_{\textrm{zpf}})^2}{l-x +x_{\textrm{int}}}\frac{TT_m/4}{(T_s +TT_m/4)^2}
\end{equation}
for the efficiency-weighted cooperativity of  the $\omega_\mathrm{R}$ mode.
It is seen that both expressions can be maximized by changing of $T$, $T_m$, or $T_s$ to match  the internal and external loss.

For the $\omega_\mathrm{L}$ mode, the maximum is reached at $T=T_s$ yielding
\begin{equation}
\label{CLM}
C_{\eta,\,\mathrm{max}}=  \frac{4}{c\gamma_m}\frac{(\omega_c x_{\textrm{zpf}})^2}{x +x_{\textrm{int}}}\frac{1}{T_s}.
\end{equation}
while, for the $\omega_\mathrm{R}$ mode, the maximum condition reads
\begin{equation}
\label{COND}
TT_m/4=T_s
\end{equation}
such that
\begin{equation}
\label{CRM}
C_{\eta,\,\mathrm{max}}=  \frac{4}{c\gamma_m}\frac{(\omega_c x_{\textrm{zpf}})^2}{l-x +x_{\textrm{int}}}\frac{1}{T_s}.
\end{equation}

 Applying the above expression to  MIM, we find
\begin{equation}
\label{CMIMmax}
C_{\eta,\,\mathrm{max}}^{\mathrm{MIM}}= \frac{8}{c\gamma_m}\frac{(\omega_c x_{\textrm{zpf}})^2}{l }\frac{1}{T_s}
\end{equation}
As for MAK and MATE, if we denote the separation between the membrane and the nearest mirror as $\delta$, Eqs. (\ref{CLM}) and  (\ref{CRM})  yields the same result
\begin{equation}
\label{CMATEmax}
C_{\eta,\,\mathrm{max}}^{\mathrm{MATE,\,MAK}}=  \frac{4}{c\gamma_m}\frac{(\omega_c x_{\textrm{zpf}})^2}{\delta +x_{\textrm{int}}}\frac{1}{T_s}.
\end{equation}
Here, the following reservation is needed. 
For highly reflecting membranes, condition (\ref{COND}) may not be met such that the efficiency-weighted cooperativity of MAK may not be optimized up to the value given by Eq.(\ref{CMATEmax}).   

Comparing Eqs. (\ref{CMIMmax}) with (\ref{CMATEmax}),  since for MAK and MATE, as agreed, $\delta\ll l$, one clearly sees that, w.r.t. the optimized efficiency-weighted  cooperativity, MATE and MAK are better than MIM.

\section{ MAK and MATE vs short Fabry-Perot cavity }
In the discussion above, we have addressed the problem of the optimal placing of a highly reflecting membrane in a one-sided cavity.
We did it for the points of the spectrum, which we called r-points, where $g_0$ is less affected by the effect of the avoided crossing of the resonance modes of the two subcavities.
It was found that MIM is always less effective than MAK and MATE.
A remarkable feature of MAK and MATE at  the r-points is that the shorter subcavity is on resonance  while the longer one is on antiresonance.
Actually, MATE can be formally viewed as a short cavity with a synthetic backstop mirror while MAK as that with a  synthetic coupling  mirror, however with a reservation that, in the regime of interest, the energy of the modes used is not mainly stored in the shorter subcavity.
In this context, it is of interest to compare the performance of MAK and MATE, which have membrane/mirror separation $\delta$, with the performance of a $\delta$  long one-sided Fabry-Perot (FP) cavity with a membrane used as the coupling mirror.

One readily finds the following principle parameters of  FP
\begin{equation}
\label{gFP}
    g_{0}^{\textrm{FP}} = \frac{\omega_c}{\delta} x_\textrm{zpf}
\end{equation}
\begin{equation}
\label{kapeFP}
\kappa_{\mathrm{ext},}^{\textrm{FP}} = \frac{cT_m}{2\delta},\qquad \kappa_{s}^{\textrm{FP}} = \frac{cT_s}{2\delta}.
\end{equation}
\begin{equation}
\label{CFP}
C^{\textrm{FP}}= \frac{8}{c\gamma_m}\frac{(\omega_c x_{\textrm{zpf}})^2}{\delta}\frac{1}{T_s +T_m}.
\end{equation}
\begin{equation}
\label{CFPet}
C_\eta^{\mathrm{FP}}= \frac{8}{c\gamma_m}\frac{(\omega_c x_{\textrm{zpf}})^2}{\delta}\frac{T_m}{(T_s +T_m)^2}
\end{equation}
\begin{equation}
\label{CFPmax}
C_{\eta,\,\mathrm{max}}^{\textrm{FP}}= \frac{4}{c\gamma_m}\frac{(\omega_c x_{\textrm{zpf}})^2}{\delta}\frac{1}{T_s}
\end{equation}

Comparing $g_0$, i.e (\ref{L}), (\ref{R}), and (\ref{gFP}), one finds that FP is advantageous by a factor of $\frac{x_{\mathrm{int}}+\delta}{\delta}$.
The same factor controls the superiority of FP in the case of the optimized efficiency-weighted  cooperativity (c.f. (\ref{CFPmax}) and  (\ref{CMATEmax})).
\begin{figure}
\includegraphics [width=0.7\columnwidth,clip=true, trim=0mm 0mm 0mm 0mm] {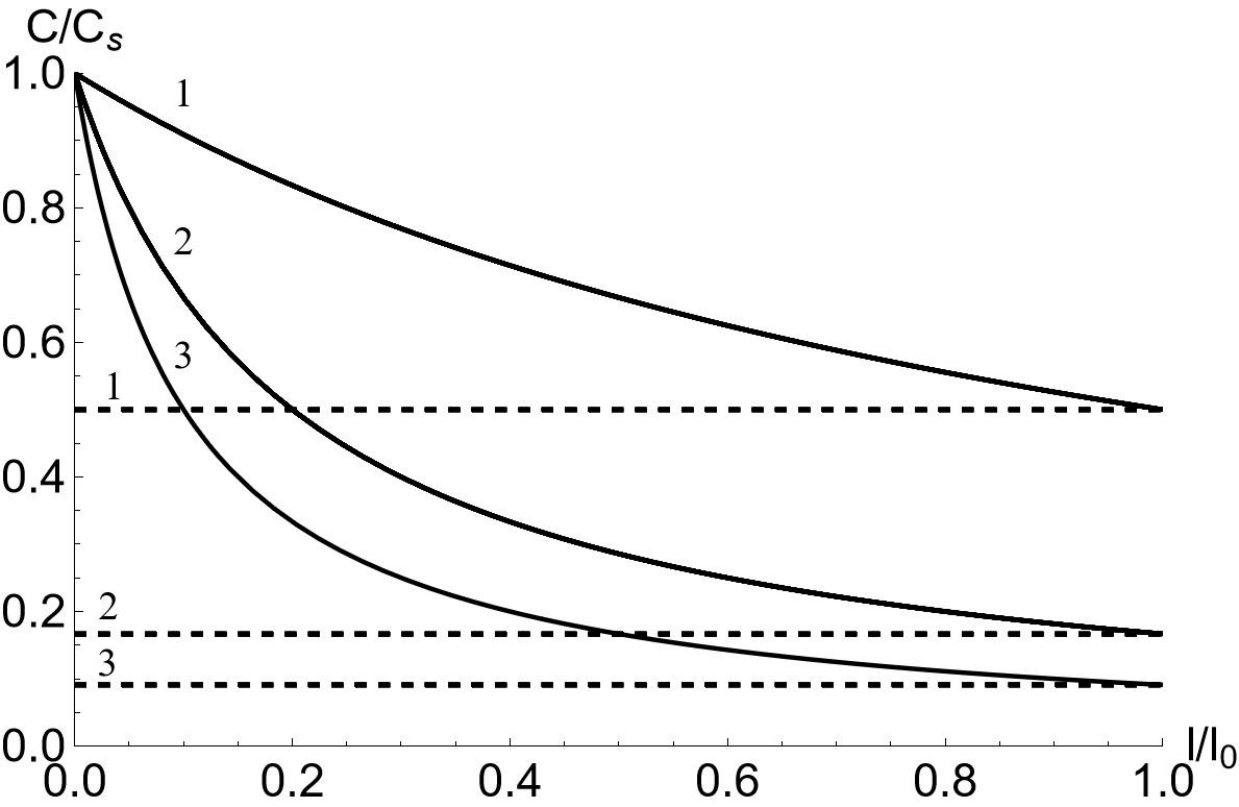}
\caption{ The normalized cooperativity $ C_{\textrm{FP}}/C_{s}$ of FP (dashed lines) and   $C_{\textrm{MAK}}/C_{s}$ of MAK (solid lines),  plotted as a function of  the normalized length of the long cavity $l/l_0$,  where $l_0=4\delta /T_s$ while  $C_{s}=8(\omega_c x_{\textrm{zpf}})^2/(c\gamma_m \delta T_s)$ is the cooperativity of  FP in the limit where the loss is fully dominated by the parasitic scattering. Numbers 1, 2, and 3 mark the curves plotted for $T_m/T_s= $2, 5, and 10, respectively.
The situation of practical interest where $T_s\gg TT_m/4$ is addressed.
\label{FF6}}
\end{figure}

For the cooperativity, we will compare FP with MAK, which, w.r.t.  this parameter,  is the best of the configurations with the membrane inside the cavity.
Thus, we will compare (\ref{CFP}) with (\ref{CR}) rewritten as follows
\begin{equation}
\label{CR1}
C^{\textrm{MAK}}=  \frac{8}{c\gamma_m}\frac{(\omega_c x_{\textrm{zpf}})^2}{\delta +lT_m/4}\frac{1}{T_s +TT_m/4}.
\end{equation}

First,  we address the regime of extremely low parasitic scattering where $T_s$ is smaller or about  $TT_m/4$ , in this regime, MAK is strongly advantageous compared to other configurations with a highly reflecting membrane inside the cavity.
One readily checks that, for not too long cavities, that is for
\begin{equation}
\label{ll}
l\ll  \frac{4\delta}{T_m},
\end{equation}
MAK is advantageous over FP by a  factor of $\frac{4(T_m+T_s)}{T_mT}$, which can make a few order-of-magnitude gain.

Such a regime, however,  is not realistic for currently available systems with highly reflecting membranes, where  $T_s\gg TT_m/4$.
At $T_s\gg TT_m/4$, (\ref{CFP}) and  (\ref{CR1})  can be simplified to find

\begin{equation}
\label{CFP1}
C^{\textrm{FP}}= \frac{C_s}{1+\frac{T_m}{T_s}},
\end{equation}
\begin{equation}
\label{CMAK}
C^{\textrm{MAK}}= \frac{C_s}{1+\frac{l}{\delta}\frac{T_m}{4}}
\end{equation}
where
\begin{equation}
\label{Cs}
C_{s}= \frac{8}{c\gamma_m}\frac{(\omega_c x_{\textrm{zpf}})^2}{\delta T_s}.
\end{equation}
is the cooperativity of  FP in the limit where the loss is fully dominated by the parasitic scattering.
Comparing (\ref{CFP1}) with (\ref{CMAK}) we find that, in cavities shorter that $l_0=4\delta /T_s$, MAK is  better than FP.
However that advantage can be appreciable only if $T_s\ll T_m$.
A comparison between the systems is also illustrated in Fig.~\ref{FF6}.
It is worth  recalling that throughout this paper we consider MAK at the r-points such that, for MAK, this figure applies only for $\delta$  and $l$ corresponding to the r-point condition, i.e. when the shorter sub-cavity  is on resonance while the longer one on  antiresonance.
\section{Optimization of an optomechanical setup}
When discussing above the advantages and disadvantages of various configuration containing a highly reflecting membrane we felt free with manipulating with all parameters of the systems.
However, in practice, the experimenter will seek to obtain the highest figure of merit for their application with a given high reflecting membrane.
Let us fix the parasitic loss $T_s$, and the minimum membrane transmission $T_{m, min}$, while allow the membrane transmission be widely tunable above this value via changing the operating wavelength.
The other quantities, which are in the hands of the experimenter, are as follows: the separation $\delta$  between the membrane and the adjacent mirror, the overall length of the cavity $l$, and the coupling mirror transmission $T$.
 All schemes favour a small $\delta$ .
Thus it should be set minimal.
 A membrane-mirror separation down to $(1.6 \pm 0.8)~\mu\text{m}$ was reported in \cite{dumont_flexure-tuned_2019}.
 Once the achievable $\delta$ has been chosen, we can discuss optimal choices for the remaining free parameters.

 It was shown above that w.r.t. the cooperativity, the best configurations are MAK or FP while  w.r.t the weighted cooperativity FP is always the best.
 For this reason, we will address the optimization  of MAK or FP  w.r.t.  the cooperativity and PF w.r.t the weighted cooperativity.

 Starting from the weighted cooperativity of  FP, via Eq.~(\ref{CFPet}) and (\ref{CFPmax}), one readily finds that it is maximal when the external loss matches the internal loss, i.e. at $T_m=T_s$.
 If, however, $ T_{m, min}>T_s$, to maximize the weighted cooperativity, one should set $ T_m=T_{m, min}$.

 When discussing the optimisation w.r.t. the cooperativity we restrict ourselves to the realistic situation where   $T_s \gg TT_m/4$ such that Eqs. (\ref{CFP1}) and (\ref{CMAK}) and Fig. \ref{FF6} can be used.
First, we see that to maximize the gain from the use of highly reflecting membrane one should use MAK with the cavity shorter than $l_0=4\delta /T_s$.
Second, as it follows from (\ref{CMAK}), for a fixed cavity length, a  further maximization of the cooperativity is possible by a reduction of the membrane transparency $T_m$.
One can also readily check that  the maximal gain of MAK over FP  is $T_m/ T_s$.
Since  $T_s$ can be as small as  $10^{-3}-10^{-4}$  \cite{chen2017high,enzian2023phononically,zhou_2023_cavity}, such a gain still can be appreciable.

\section{Conclusions}

We present an analysis of  a problem of the optimal position of a highly reflecting membrane in a one-sided cavity.
In our analysis, we addressed the coupling constant and the figures of merit of an opto-mechanical device such as the cooperativity and  efficiency-weighted cooperativity.
These figures of merit are relevant in different situations, e.g. for  optomechanical cooling the cooperativity matters while it is the efficiency-weighted cooperativity that matters for optical sensing of  the mechanical subsystem.
It was found that the optimal settings for these figures of merit are very different.

In contrast to the previous  theoretical considerations of an optical cavity with a membrane inside \cite{jayich_dispersive_2008, miao_standard_2009,dumont_flexure-tuned_2019} here we have incorporated the parasitic scattering of the membrane  while neglecting the parasitic scattering from the coupling mirror, which we assume to be much weaker than that from the membrane.
We have demonstrated that, in the case of a highly reflecting membrane, even a small amount of parasitic scattering from the membrane may have an essential impact on the cooperativity of the system.

The regimes with the membrane close to the coupling mirror (MATE), close to the backstop mirror (MAK), and close to the cavity center (MIM) were compared.
The comparison was done for the points of the spectrum, which we called r-points, where $g_0$ is less affected by the effect of the avoided crossing of the resonance modes of the two subcavities.
A remarkable feature of MAK and MATE at  the r-points is that the shorter subcavity is on resonance  while the longer one  is on antiresonance such that  MATE can be viewed as a short cavity with a synthetic backstop mirror while MAK as that with a  synthetic coupling  mirror.
In this context,  we compared the optomechanical parameters  of MATE and  MAK with those of a  one-sided Fabry-Perot cavity (FP) with a membrane used as the coupling mirror,  its length $\delta$ being equal  to the separation between the membrane and the adjacent mirror in MAK or MATE.

It has been found that, w.r.t.  the coupling constant and the efficiency-weighted  cooperativity, FP is the best of all the systems addressed.
However, in terms of the cooperativity, the situation of different.
Among   MIM, MATE and  MAK, the latter is always the best though,  in some regimes, the performance of MATE  can  be very close  to that of MAK.
Comparing  MAK  with FP, it has been found that each of them can be superior, depending on the parameters of the system.
First, in the limit of the very weak parasitic scattering, the regime where $T_s\ll TT_m/4$,  MAK  is  superior over  FP by a factor of about $\frac{4(T_m+T_s)}{T_mT}$.
Thus, in this regime,  MAK would provide an extraordinary performance.
However, even a very small amount of parasitic scattering from the membrane can essentially suppress the performance of  MAK.
This happens when $T_s$ is comparable or larger than $TT_m/4$.
The situation of practical interest is $T_s\gg TT_m/4$.
In this regime,  FP can compete with MAK, specifically, for  the cavity length $l$ used in MAK exceeding  $l_0=4\delta /T_s$ , the cooperativity of FP is larger than that of MAK.
For shorter cavities, the cooperativity of MAK is larger than that of FP.
For the  membranes having the parasitic scattering much weaker than the transmission, i.e. for $T_s/T_m\ll 1$ and $l \ll4\delta /T_s$, this advantage is appreciable being about  $ T_m/T_s$.

All in all, in terms of cooperativity, for the opto-mechanical setups using a highly reflecting membrane placed inside an optical cavity, the MAK configuration, is shown to be advantageous compared to other configurations.
However, in this respect, in a certain regime, the performance of MATE can be very close to that of MAK.
In addition,  in terms of cooperativity, MAK can also be advantageous compared to FP with the membrane as the coupling mirror with the same membrane/mirror separation as MAK.
 In the regime of extremely weak parasitic scattering of the membrane,  MAK can provide very high cooperativity.
 Outside of this regime, which currently is a realistic experimental situation, MAK can be tuned to yield a cooperativity larger than that of FP.
 This advantage can be essential only for a membrane exhibiting the  parasitic scattering appreciably weaker than  its transmission.
 If one cares about the efficiency-weighted  cooperativity, the use of a highly reflecting membrane is optimal when it serves as the coupling mirror of a FP cavity.

\section{Funding}
G.E. acknowledges support from the European Union’s Horizon 2020 research and innovation programme under the Marie Sklodowska-Curie grant agreement No. 847523.
E.S.P has been supported by VILLUM FONDEN under a Villum Investigator Grant, no. 25880.
\section{Acknowledgments}
The authors acknowledge insightful  discussions with A. Simonsen and Z. Wang.

\section{Disclosures}
The authors declare no conflicts of interest.

\section{Data availability} Data underlying the results presented in this paper are not publicly available at this time but may be obtained from the authors upon reasonable request.

\appendix
\section{Frequency gap at the crossing points}
\label{Gap}
In this Appendix, we evaluate the frequency gap which appears at a crossing point of the spectrum of the  sub-cavities of the systems (see Fig.\ref{FF1}) once a weak coupling between these is allowed.  
A form of equation~(\ref{fr}) that is proper for further analysis reads
\begin{equation}
\label{fr1}
\cos[k(l-x)+\varphi_r/2]\cos(kx+\varphi_r/2) =  \frac{t_m^2}{4}\cos(2kx-kl).
\end{equation}
Let $x_0$ and $k_0$ be the membrane position and the resonant wave-number corresponding to a crossing point, respectively.
At these points, both sub-cavities  taken decoupled are on resonance such that both cosines in (\ref{fr1}) should be equal to zero:
\begin{equation}
\label{set}
\cos[k_0(l-x_0)+\varphi_r/2]=0\qquad \textrm{and}\qquad \cos(k_0x_0+\varphi_r/2) =0.
\end{equation}
Next, looking for the solution to equation (\ref{fr1}) in the form $x=x_0$ and  $k=k_0+\delta k$, we find, in the limit of small $t_m$, that
\begin{equation}
\label{gap1}
\delta k^2\propto  \frac{t_m^2}{x_0(l-x_0)},
\end{equation}
implying (\ref{gap0}).
\section{Properties of the system at  the r-points}
\label{rpoints}
This Appendix addresses some properties of the system at  the r-points.

Let us show that, with a variation of $t_m$, the $\omega(x)$ curves "locally rotate" about the r-points.
To be specific let us consider the r-points formed by the intersection of the $\omega(x)$ curves for the $\omega_\mathrm{L}$ mode calculated at $t_m=0$ and $t_m\neq0$.
Since we are on the resonance curve for the $x$-long part of the cavity taken decoupled from the other part, the phase shift on reflection from the membrane equals $\varphi_r$.
Thus, taking into account that  the round-trip phase variation along any loop should be equal to $2\pi$ times an integer, we can write
\begin{equation}
\label{Lresonance}
e^{i(2kx +\pi+\varphi_r)}=1.
\end{equation}
For the $(l-x)$-long part, we can also write the following round-trip phase condition
\begin{equation}
\label{Rresonance}
e^{i[2k(l-x) +\pi+\mu]}=1,
\end{equation}
where $\mu$ is the phase shift at the reflection from the $x$-long part.
Since the $(l-x)$-long part taken decoupled is not on resonance $\mu\neq \varphi_r$.
These two equations specify the positions of the corresponding r-points.
On the other hand, one readily checks that the phase shift  on reflection from a one-sided cavity on resonance is independent of transmission of the input mirror (see Appendix \ref{Reflection}).
For our setting
\begin{equation}
\label{mu}
\mu=\varphi_r -\pi.
\end{equation}
Thus, $\mu$ is independent of the membrane transmission, implying the independence of positions of the r-points.

For  the $\omega_\mathrm{L}$ modes, Eqs. (\ref{Lresonance}), (\ref{Rresonance}), and (\ref{mu}) can be rewritten as follows 
\begin{equation}
\label{resL0}
e^{-2ik(l-x) -i\varphi_r}=1,\qquad  e^{-2ikx -i\varphi_r}=-1.
\end{equation}
Similar  expressions can be readily shown  for the $\omega_\mathrm{R}$ modes:
\begin{equation}
\label{resR0}
e^{-2ik(l-x) -i\varphi_r}=-1,\qquad  e^{-2ikx -i\varphi_r}=1.
\end{equation}
Equations (\ref{resL0}) and (\ref{resR0}) imply that, at the the r-points,
one of the sub-cavities is on resonance while the other is on antiresonance .

For the r-points, we also evaluate the ratio of the field intensity in one part of the cavity to that in the other part.
We will first consider this for an $\omega_\mathrm{L}$ mode.
Basing on equation (\ref{se1}), we consider the following complex amplitude balance equation at the membrane (compare Fig. \ref{FF1})
\begin{equation}
\label{balance}
B=t_me^{i\varphi_s}A+r_me^{i\varphi_r}G \ .
\end{equation}
Since, at the r-points considered, the x-long part taken decoupled is on resonance, using (\ref{mu}), we can write
\begin{equation}
\label{balance1}
B=e^{i(\varphi_r-\pi)}G. 
\end{equation}
Combining equations (\ref{balance}) and (\ref{balance1}), in the limit $t_m\ll1$, we find
\begin{equation}
\label{ratio00}
\frac{|B|^2}{|A|^2}=\frac{T_m}{4},
\end{equation}
where $T_m=t_m^2$ is the  power (intensity) transmission of the membrane while $|A|^2$ and $|B|^2$ are the  intensities in the x-long and l-x-long parts of the cavity, respectively.

For the r-points of the $\omega_\mathrm{R}$ mode, similar calculations yield
\begin{equation}
\label{ratio11}
\frac{|B|^2}{|A|^2}=\frac{4}{T_m}.
\end{equation}

Next, we find the mirror/membrane separation,  labeled as  $x_\textrm{int}$, at which, for the r-points, the energy of a mode  stored in one sub-cavity is equal to that stored in the other.
Evidently, $x_\textrm{int}$ is given by the solution to the following equation 
\begin{equation}
\label{half}
x|A|^2=(l-x)|B|^2
\end{equation}
such that, for a highly reflecting membrane, using (\ref{ratio00}) and (\ref{ratio11}), we find
\begin{equation}
\label{half1}
x_{\textrm{int}}=l\frac{T_m}{4}.
\end{equation}

In application to the $\omega_\mathrm{L}$ modes, (\ref{half1}) gives the membrane separation from the coupling mirror while, for the $\omega_\mathrm{R}$ modes, it is the separation from the backstop mirror.
\section{Reflection from resonance cavity }
\label{Reflection}
Consider only the membrane and the backstop mirror as in  Fig.\ref{FF1}, when the $l-x$ long part of the cavity is on resonance, i.e. according to (\ref{resR})
\begin{equation}
\label{resR1}
e^{-2ik_0(l-x) -i\varphi_r}=-1.
\end{equation}
According to (\ref{mirrorB}), (\ref{membrane}) and (\ref{parameters}) the complex amplitudes are linked by the following relations
\begin{align}
\begin{aligned}
\\&B=-Ce^{-2ik(l-x)},
\\&B=t_me^{i\varphi_t}A + r_me^{i\varphi_r} G,
\\&F=r_me^{i\varphi_r} A+t_me^{i\varphi_t}G,
\end{aligned}
\label{se1}
\end{align}
Eliminating $B$ and $C$ between set (\ref{se1}) and taking into account (\ref{resR1}) we find
\begin{equation}
\label{refl}
\frac{F}{A}= -e^{i\varphi_r}
\end{equation}
implying that the phase of the signal reflected from a cavity on resonance is independent of  the modulus of the reflection coefficient of the coupling mirror.
\section{Coupling constant and decay rate}
\label{fullset}
Equations  (\ref{mirror}), (\ref{mirrorB}), (\ref{membrane}) and (\ref{parameters}) lead to the following relations between the complex amplitudes (see Fig.~\ref{FF1})
\begin{align}
\begin{aligned}
\\&A=-rFe^{2ikx},
\\&B=-Ge^{-2ik(l-x)},
\\&B=t_me^{i\varphi_t}A + r_me^{i\varphi_r} G,
\\&F=r_me^{i\varphi_r} A+t_me^{i\varphi_t}G,
\end{aligned}
\label{set2}
\end{align}
which imply the following  equation for the resonance wave vector (in general complex)
\begin{equation}
\label{resFUL}
(r_m +e^{-2ik(l-x) -i\varphi_r})(r_m +r^{-1}e^{-2ikx -i\varphi_r})+t_m^2 = 0.
\end{equation}
Since the dissipation is assumed to be weak, to find $g_0$, we neglect it by setting $r=1$ and $ t_m^2=1-r_m^2$ to find
\begin{equation}
\label{resRE}
(r_m +e^{-2ik(l-x) -i\varphi_r})(r_m +e^{-2ikx -i\varphi_r})+1-r_m^2 = 0,
\end{equation}
which is equivalent to (\ref{fr}).
The derivative $dk/dx$ calculated on the resonance using (\ref{resRE}) reads
\begin{equation}
\label{deriv}
\frac{dk}{dx}=\frac{\omega_c}{c} \frac{e^{-2ik(l-x) -i\varphi_r}(r_m +e^{-2ikx -i\varphi_r})- e^{-2ikx -i\varphi_r}(r_m +e^{-2ik(l-x) -i\varphi_r})}
{(l-x)e^{-2ik(l-x) -i\varphi_r}(r_m +e^{-2ikx -i\varphi_r})+x e^{-2ikx -i\varphi_r}(r_m +e^{-2ik(l-x) -i\varphi_r})}.
\end{equation}
For the r-points of  the $\omega_\mathrm{R}$ mode
where the $l-x$ long subcavity is on resonance while the other subcavity is on  antiresonance. i.e.  $e^{-2ik_0(l-x) -i\varphi_r}=-1$ and  $ e^{-2ik_0x -i\varphi_r}=1$,
we find
\begin{equation}
\label{derivR}
\frac{dk}{dx}=\frac{\omega_c}{c} \frac{r_m}
{(l-x)r_m+lt_m^2/4},
\end{equation}
which,  in the approximation $t_m^2\ll 1$, readily  yields (\ref{R}) .

For the r-points of the $\omega_\mathrm{L}$ modes
where $e^{-2ik_0(l-x) -i\varphi_r}=1$ and  $ e^{-2ik_0x -i\varphi_r}=-1$,
on the same lines, we obtain  (\ref{L}) as well.

To evaluate the cavity decay rate, we rewrite (\ref{resFUL}) as follows
\begin{equation}
\label{resDISS0}
(r_m +e^{-2ik(l-x) -i\varphi_r})(r_m +r^{-1}e^{-2ikx -i\varphi_r})+1- r_m^2 = T_s.
\end{equation}
We expand (\ref{resDISS0}) about $k_0$, which is the solution to (\ref{resRE}),  keeping the lowest term in $t^2\ll 1$ and  $\delta k =k-k_0$, we find
\begin{equation}
\label{deltak}
\delta k=-\frac{i}{2} \frac{T_s+ (1-r^{-1})e^{-2ikx -i\varphi_r}(r_m +e^{-2ik(l-x) -i\varphi_r})}
{(l-x)e^{-2ik(l-x) -i\varphi_r}(r_m +e^{-2ikx -i\varphi_r})+x e^{-2ikx -i\varphi_r}(r_m +e^{-2ik(l-x) -i\varphi_r})}.
\end{equation}
Now, using the above resonance/antiresonace  conditions for the sub-cavities, for  the $\omega_\mathrm{R}$ mode, we arrive at the following cavity decay rate:
\begin{equation}
\label{decayR}
\kappa=-2c\mathrm{Im}[\delta k] =\frac{c}{2} \frac{T_S+(r^{-1}-1)(1-r_m)}
{(l-x)r_m+lt_m^2/4},
\end{equation}
 which, under conditions $1-r_m\approx T_m/2\ll 1$ and $1-r\approx T/2\ll 1$, readily brings us to (\ref{dampingRe}) and (\ref{dampingRs}).

 On the same lines one obtains (\ref{dampingLe}) and (\ref{dampingLs}).
\section*{References}

\bibliography{LIT}

\begin{thebibliography}{15}%
\makeatletter
\providecommand \@ifxundefined [1]{%
 \@ifx{#1\undefined}
}%
\providecommand \@ifnum [1]{%
 \ifnum #1\expandafter \@firstoftwo
 \else \expandafter \@secondoftwo
 \fi
}%
\providecommand \@ifx [1]{%
 \ifx #1\expandafter \@firstoftwo
 \else \expandafter \@secondoftwo
 \fi
}%
\providecommand \natexlab [1]{#1}%
\providecommand \enquote  [1]{``#1''}%
\providecommand \bibnamefont  [1]{#1}%
\providecommand \bibfnamefont [1]{#1}%
\providecommand \citenamefont [1]{#1}%
\providecommand \href@noop [0]{\@secondoftwo}%
\providecommand \href [0]{\begingroup \@sanitize@url \@href}%
\providecommand \@href[1]{\@@startlink{#1}\@@href}%
\providecommand \@@href[1]{\endgroup#1\@@endlink}%
\providecommand \@sanitize@url [0]{\catcode `\\12\catcode `\$12\catcode
  `\&12\catcode `\#12\catcode `\^12\catcode `\_12\catcode `\%12\relax}%
\providecommand \@@startlink[1]{}%
\providecommand \@@endlink[0]{}%
\providecommand \url  [0]{\begingroup\@sanitize@url \@url }%
\providecommand \@url [1]{\endgroup\@href {#1}{\urlprefix }}%
\providecommand \urlprefix  [0]{URL }%
\providecommand \Eprint [0]{\href }%
\providecommand \doibase [0]{http://dx.doi.org/}%
\providecommand \selectlanguage [0]{\@gobble}%
\providecommand \bibinfo  [0]{\@secondoftwo}%
\providecommand \bibfield  [0]{\@secondoftwo}%
\providecommand \translation [1]{[#1]}%
\providecommand \BibitemOpen [0]{}%
\providecommand \bibitemStop [0]{}%
\providecommand \bibitemNoStop [0]{.\EOS\space}%
\providecommand \EOS [0]{\spacefactor3000\relax}%
\providecommand \BibitemShut  [1]{\csname bibitem#1\endcsname}%
\let\auto@bib@innerbib\@empty
\bibitem [{\citenamefont {Jayich}\ \emph {et~al.}(2008)\citenamefont {Jayich},
  \citenamefont {Sankey}, \citenamefont {Zwickl}, \citenamefont {Yang},
  \citenamefont {Thompson}, \citenamefont {Girvin}, \citenamefont {Clerk},
  \citenamefont {Marquardt},\ and\ \citenamefont
  {Harris}}]{jayich_dispersive_2008}%
  \BibitemOpen
  \bibfield  {author} {\bibinfo {author} {\bibfnamefont {A.~M.}\ \bibnamefont
  {Jayich}}, \bibinfo {author} {\bibfnamefont {J.~C.}\ \bibnamefont {Sankey}},
  \bibinfo {author} {\bibfnamefont {B.~M.}\ \bibnamefont {Zwickl}}, \bibinfo
  {author} {\bibfnamefont {C.}~\bibnamefont {Yang}}, \bibinfo {author}
  {\bibfnamefont {J.~D.}\ \bibnamefont {Thompson}}, \bibinfo {author}
  {\bibfnamefont {S.~M.}\ \bibnamefont {Girvin}}, \bibinfo {author}
  {\bibfnamefont {A.~A.}\ \bibnamefont {Clerk}}, \bibinfo {author}
  {\bibfnamefont {F.}~\bibnamefont {Marquardt}}, \ and\ \bibinfo {author}
  {\bibfnamefont {J.~G.~E.}\ \bibnamefont {Harris}},\ }\href {\doibase
  10.1088/1367-2630/10/9/095008} {\bibfield  {journal} {\bibinfo  {journal}
  {New Journal of Physics}\ }\textbf {\bibinfo {volume} {10}},\ \bibinfo
  {pages} {095008} (\bibinfo {year} {2008})}\BibitemShut {NoStop}%
\bibitem [{\citenamefont {Miao}\ \emph {et~al.}(2009)\citenamefont {Miao},
  \citenamefont {Danilishin}, \citenamefont {Corbitt},\ and\ \citenamefont
  {Chen}}]{miao_standard_2009}%
  \BibitemOpen
  \bibfield  {author} {\bibinfo {author} {\bibfnamefont {H.}~\bibnamefont
  {Miao}}, \bibinfo {author} {\bibfnamefont {S.}~\bibnamefont {Danilishin}},
  \bibinfo {author} {\bibfnamefont {T.}~\bibnamefont {Corbitt}}, \ and\
  \bibinfo {author} {\bibfnamefont {Y.}~\bibnamefont {Chen}},\ }\href {\doibase
  10.1103/PhysRevLett.103.100402} {\bibfield  {journal} {\bibinfo  {journal}
  {Physical Review Letters}\ }\textbf {\bibinfo {volume} {103}},\ \bibinfo
  {pages} {100402} (\bibinfo {year} {2009})},\ \bibinfo {note} {publisher:
  American Physical Society}\BibitemShut {NoStop}%
\bibitem [{\citenamefont {Yanay}\ \emph {et~al.}(2016)\citenamefont {Yanay},
  \citenamefont {Sankey},\ and\ \citenamefont {Clerk}}]{yanay_quantum_2016}%
  \BibitemOpen
  \bibfield  {author} {\bibinfo {author} {\bibfnamefont {Y.}~\bibnamefont
  {Yanay}}, \bibinfo {author} {\bibfnamefont {J.~C.}\ \bibnamefont {Sankey}}, \
  and\ \bibinfo {author} {\bibfnamefont {A.~A.}\ \bibnamefont {Clerk}},\ }\href
  {\doibase 10.1103/PhysRevA.93.063809} {\bibfield  {journal} {\bibinfo
  {journal} {Physical Review A}\ }\textbf {\bibinfo {volume} {93}},\ \bibinfo
  {pages} {063809} (\bibinfo {year} {2016})},\ \bibinfo {note} {publisher:
  American Physical Society}\BibitemShut {NoStop}%
\bibitem [{\citenamefont {Thompson}\ \emph {et~al.}(2008)\citenamefont
  {Thompson}, \citenamefont {Zwickl}, \citenamefont {Jayich}, \citenamefont
  {Marquardt}, \citenamefont {Girvin},\ and\ \citenamefont
  {Harris}}]{thompson_strong_2008}%
  \BibitemOpen
  \bibfield  {author} {\bibinfo {author} {\bibfnamefont {J.~D.}\ \bibnamefont
  {Thompson}}, \bibinfo {author} {\bibfnamefont {B.~M.}\ \bibnamefont
  {Zwickl}}, \bibinfo {author} {\bibfnamefont {A.~M.}\ \bibnamefont {Jayich}},
  \bibinfo {author} {\bibfnamefont {F.}~\bibnamefont {Marquardt}}, \bibinfo
  {author} {\bibfnamefont {S.~M.}\ \bibnamefont {Girvin}}, \ and\ \bibinfo
  {author} {\bibfnamefont {J.~G.~E.}\ \bibnamefont {Harris}},\ }\href {\doibase
  10.1038/nature06715} {\bibfield  {journal} {\bibinfo  {journal} {Nature}\
  }\textbf {\bibinfo {volume} {452}},\ \bibinfo {pages} {72} (\bibinfo {year}
  {2008})}\BibitemShut {NoStop}%
\bibitem [{\citenamefont {Wilson}\ \emph {et~al.}(2009)\citenamefont {Wilson},
  \citenamefont {Regal}, \citenamefont {Papp},\ and\ \citenamefont
  {Kimble}}]{wilson_cavity_2009}%
  \BibitemOpen
  \bibfield  {author} {\bibinfo {author} {\bibfnamefont {D.~J.}\ \bibnamefont
  {Wilson}}, \bibinfo {author} {\bibfnamefont {C.~A.}\ \bibnamefont {Regal}},
  \bibinfo {author} {\bibfnamefont {S.~B.}\ \bibnamefont {Papp}}, \ and\
  \bibinfo {author} {\bibfnamefont {H.~J.}\ \bibnamefont {Kimble}},\ }\href
  {\doibase 10.1103/PhysRevLett.103.207204} {\bibfield  {journal} {\bibinfo
  {journal} {Physical Review Letters}\ }\textbf {\bibinfo {volume} {103}},\
  \bibinfo {pages} {207204} (\bibinfo {year} {2009})}\BibitemShut {NoStop}%
\bibitem [{\citenamefont {Purdy}\ \emph {et~al.}(2013)\citenamefont {Purdy},
  \citenamefont {Yu}, \citenamefont {Peterson}, \citenamefont {Kampel},\ and\
  \citenamefont {Regal}}]{purdy_strong_2013}%
  \BibitemOpen
  \bibfield  {author} {\bibinfo {author} {\bibfnamefont {T.~P.}\ \bibnamefont
  {Purdy}}, \bibinfo {author} {\bibfnamefont {P.-L.}\ \bibnamefont {Yu}},
  \bibinfo {author} {\bibfnamefont {R.~W.}\ \bibnamefont {Peterson}}, \bibinfo
  {author} {\bibfnamefont {N.~S.}\ \bibnamefont {Kampel}}, \ and\ \bibinfo
  {author} {\bibfnamefont {C.~A.}\ \bibnamefont {Regal}},\ }\href {\doibase
  10.1103/PhysRevX.3.031012} {\bibfield  {journal} {\bibinfo  {journal}
  {Physical Review X}\ }\textbf {\bibinfo {volume} {3}},\ \bibinfo {pages}
  {031012} (\bibinfo {year} {2013})}\BibitemShut {NoStop}%
\bibitem [{\citenamefont {Mason}\ \emph {et~al.}(2019)\citenamefont {Mason},
  \citenamefont {Chen}, \citenamefont {Rossi}, \citenamefont {Tsaturyan},\ and\
  \citenamefont {Schliesser}}]{mason_continuous_2019}%
  \BibitemOpen
  \bibfield  {author} {\bibinfo {author} {\bibfnamefont {D.}~\bibnamefont
  {Mason}}, \bibinfo {author} {\bibfnamefont {J.}~\bibnamefont {Chen}},
  \bibinfo {author} {\bibfnamefont {M.}~\bibnamefont {Rossi}}, \bibinfo
  {author} {\bibfnamefont {Y.}~\bibnamefont {Tsaturyan}}, \ and\ \bibinfo
  {author} {\bibfnamefont {A.}~\bibnamefont {Schliesser}},\ }\href {\doibase
  10.1038/s41567-019-0533-5} {\bibfield  {journal} {\bibinfo  {journal} {Nature
  Physics}\ }\textbf {\bibinfo {volume} {15}},\ \bibinfo {pages} {745}
  (\bibinfo {year} {2019})}\BibitemShut {NoStop}%
\bibitem [{\citenamefont {Kampel}\ \emph {et~al.}(2017)\citenamefont {Kampel},
  \citenamefont {Peterson}, \citenamefont {Fischer}, \citenamefont {Yu},
  \citenamefont {Cicak}, \citenamefont {Simmonds}, \citenamefont {Lehnert},\
  and\ \citenamefont {Regal}}]{kampel_improving_2017}%
  \BibitemOpen
  \bibfield  {author} {\bibinfo {author} {\bibfnamefont {N.}~\bibnamefont
  {Kampel}}, \bibinfo {author} {\bibfnamefont {R.}~\bibnamefont {Peterson}},
  \bibinfo {author} {\bibfnamefont {R.}~\bibnamefont {Fischer}}, \bibinfo
  {author} {\bibfnamefont {P.-L.}\ \bibnamefont {Yu}}, \bibinfo {author}
  {\bibfnamefont {K.}~\bibnamefont {Cicak}}, \bibinfo {author} {\bibfnamefont
  {R.}~\bibnamefont {Simmonds}}, \bibinfo {author} {\bibfnamefont
  {K.}~\bibnamefont {Lehnert}}, \ and\ \bibinfo {author} {\bibfnamefont
  {C.}~\bibnamefont {Regal}},\ }\href {\doibase 10.1103/PhysRevX.7.021008}
  {\bibfield  {journal} {\bibinfo  {journal} {Physical Review X}\ }\textbf
  {\bibinfo {volume} {7}},\ \bibinfo {pages} {021008} (\bibinfo {year}
  {2017})},\ \bibinfo {note} {publisher: American Physical Society}\BibitemShut
  {NoStop}%
\bibitem [{\citenamefont {Higginbotham}\ \emph {et~al.}(2018)\citenamefont
  {Higginbotham}, \citenamefont {Burns}, \citenamefont {Urmey}, \citenamefont
  {Peterson}, \citenamefont {Kampel}, \citenamefont {Brubaker}, \citenamefont
  {Smith}, \citenamefont {Lehnert},\ and\ \citenamefont
  {Regal}}]{higginbotham_harnessing_2018}%
  \BibitemOpen
  \bibfield  {author} {\bibinfo {author} {\bibfnamefont {A.~P.}\ \bibnamefont
  {Higginbotham}}, \bibinfo {author} {\bibfnamefont {P.~S.}\ \bibnamefont
  {Burns}}, \bibinfo {author} {\bibfnamefont {M.~D.}\ \bibnamefont {Urmey}},
  \bibinfo {author} {\bibfnamefont {R.~W.}\ \bibnamefont {Peterson}}, \bibinfo
  {author} {\bibfnamefont {N.~S.}\ \bibnamefont {Kampel}}, \bibinfo {author}
  {\bibfnamefont {B.~M.}\ \bibnamefont {Brubaker}}, \bibinfo {author}
  {\bibfnamefont {G.}~\bibnamefont {Smith}}, \bibinfo {author} {\bibfnamefont
  {K.~W.}\ \bibnamefont {Lehnert}}, \ and\ \bibinfo {author} {\bibfnamefont
  {C.~A.}\ \bibnamefont {Regal}},\ }\href {\doibase 10.1038/s41567-018-0210-0}
  {\bibfield  {journal} {\bibinfo  {journal} {Nature Physics}\ }\textbf
  {\bibinfo {volume} {14}},\ \bibinfo {pages} {1038} (\bibinfo {year}
  {2018})},\ \bibinfo {note} {bandiera\_abtest: a Cg\_type: Nature Research
  Journals Number: 10 Primary\_atype: Research Publisher: Nature Publishing
  Group Subject\_term: Quantum information;Quantum optics Subject\_term\_id:
  quantum-information;quantum-optics}\BibitemShut {NoStop}%
\bibitem [{\citenamefont {Dumont}\ \emph {et~al.}(2019)\citenamefont {Dumont},
  \citenamefont {Bernard}, \citenamefont {Reinhardt}, \citenamefont {Kato},
  \citenamefont {Ruf},\ and\ \citenamefont
  {Sankey}}]{dumont_flexure-tuned_2019}%
  \BibitemOpen
  \bibfield  {author} {\bibinfo {author} {\bibfnamefont {V.}~\bibnamefont
  {Dumont}}, \bibinfo {author} {\bibfnamefont {S.}~\bibnamefont {Bernard}},
  \bibinfo {author} {\bibfnamefont {C.}~\bibnamefont {Reinhardt}}, \bibinfo
  {author} {\bibfnamefont {A.}~\bibnamefont {Kato}}, \bibinfo {author}
  {\bibfnamefont {M.}~\bibnamefont {Ruf}}, \ and\ \bibinfo {author}
  {\bibfnamefont {J.~C.}\ \bibnamefont {Sankey}},\ }\href {\doibase
  10.1364/OE.27.025731} {\bibfield  {journal} {\bibinfo  {journal} {Optics
  Express}\ }\textbf {\bibinfo {volume} {27}},\ \bibinfo {pages} {25731}
  (\bibinfo {year} {2019})}\BibitemShut {NoStop}%
\bibitem [{\citenamefont {Chen}\ \emph {et~al.}(2017)\citenamefont {Chen},
  \citenamefont {Chardin}, \citenamefont {Makles}, \citenamefont {Caër},
  \citenamefont {Chua}, \citenamefont {Braive}, \citenamefont {Robert-Philip},
  \citenamefont {Briant}, \citenamefont {Cohadon}, \citenamefont {Heidmann},
  \citenamefont {Jacqmin},\ and\ \citenamefont {Deléglise}}]{chen2017high}%
  \BibitemOpen
  \bibfield  {author} {\bibinfo {author} {\bibfnamefont {X.}~\bibnamefont
  {Chen}}, \bibinfo {author} {\bibfnamefont {C.}~\bibnamefont {Chardin}},
  \bibinfo {author} {\bibfnamefont {K.}~\bibnamefont {Makles}}, \bibinfo
  {author} {\bibfnamefont {C.}~\bibnamefont {Caër}}, \bibinfo {author}
  {\bibfnamefont {S.}~\bibnamefont {Chua}}, \bibinfo {author} {\bibfnamefont
  {R.}~\bibnamefont {Braive}}, \bibinfo {author} {\bibfnamefont
  {I.}~\bibnamefont {Robert-Philip}}, \bibinfo {author} {\bibfnamefont
  {T.}~\bibnamefont {Briant}}, \bibinfo {author} {\bibfnamefont {P.-F.}\
  \bibnamefont {Cohadon}}, \bibinfo {author} {\bibfnamefont {A.}~\bibnamefont
  {Heidmann}}, \bibinfo {author} {\bibfnamefont {T.}~\bibnamefont {Jacqmin}}, \
  and\ \bibinfo {author} {\bibfnamefont {S.}~\bibnamefont {Deléglise}},\
  }\href {\doibase 10.1038/lsa.2016.190} {\bibfield  {journal} {\bibinfo
  {journal} {Light: Science \& Applications}\ }\textbf {\bibinfo {volume}
  {6}},\ \bibinfo {pages} {e16190} (\bibinfo {year} {2017})}\BibitemShut
  {NoStop}%
\bibitem [{\citenamefont {Enzian}\ \emph {et~al.}(2023)\citenamefont {Enzian},
  \citenamefont {Wang}, \citenamefont {Simonsen}, \citenamefont {Mathiassen},
  \citenamefont {Vibel}, \citenamefont {Tsaturyan}, \citenamefont {Tagantsev},
  \citenamefont {Schliesser},\ and\ \citenamefont
  {Polzik}}]{enzian2023phononically}%
  \BibitemOpen
  \bibfield  {author} {\bibinfo {author} {\bibfnamefont {G.}~\bibnamefont
  {Enzian}}, \bibinfo {author} {\bibfnamefont {Z.}~\bibnamefont {Wang}},
  \bibinfo {author} {\bibfnamefont {A.}~\bibnamefont {Simonsen}}, \bibinfo
  {author} {\bibfnamefont {J.}~\bibnamefont {Mathiassen}}, \bibinfo {author}
  {\bibfnamefont {T.}~\bibnamefont {Vibel}}, \bibinfo {author} {\bibfnamefont
  {Y.}~\bibnamefont {Tsaturyan}}, \bibinfo {author} {\bibfnamefont
  {A.}~\bibnamefont {Tagantsev}}, \bibinfo {author} {\bibfnamefont
  {A.}~\bibnamefont {Schliesser}}, \ and\ \bibinfo {author} {\bibfnamefont
  {E.~S.}\ \bibnamefont {Polzik}},\ }\href {\doibase 10.1364/OE.484369}
  {\bibfield  {journal} {\bibinfo  {journal} {Opt. Express}\ }\textbf {\bibinfo
  {volume} {31}},\ \bibinfo {pages} {13040} (\bibinfo {year}
  {2023})}\BibitemShut {NoStop}%
\bibitem [{\citenamefont {Zhou}\ \emph {et~al.}(2023)\citenamefont {Zhou},
  \citenamefont {Bao}, \citenamefont {Gorman},\ and\ \citenamefont
  {Lawall}}]{zhou_2023_cavity}%
  \BibitemOpen
  \bibfield  {author} {\bibinfo {author} {\bibfnamefont {F.}~\bibnamefont
  {Zhou}}, \bibinfo {author} {\bibfnamefont {Y.}~\bibnamefont {Bao}}, \bibinfo
  {author} {\bibfnamefont {J.~J.}\ \bibnamefont {Gorman}}, \ and\ \bibinfo
  {author} {\bibfnamefont {J.~R.}\ \bibnamefont {Lawall}},\ }\href {\doibase
  https://doi.org/10.1002/lpor.202300008} {\bibfield  {journal} {\bibinfo
  {journal} {Laser \& Photonics Reviews}\ }\textbf {\bibinfo {volume} {2023}},\
  \bibinfo {pages} {2300008} (\bibinfo {year} {2023})}\BibitemShut {NoStop}%
\bibitem [{\citenamefont {Wilson}\ \emph {et~al.}(2015)\citenamefont {Wilson},
  \citenamefont {Sudhir}, \citenamefont {Piro}, \citenamefont {Schilling},
  \citenamefont {Ghadimi},\ and\ \citenamefont
  {Kippenberg}}]{wilson_measurement-based_2015}%
  \BibitemOpen
  \bibfield  {author} {\bibinfo {author} {\bibfnamefont {D.~J.}\ \bibnamefont
  {Wilson}}, \bibinfo {author} {\bibfnamefont {V.}~\bibnamefont {Sudhir}},
  \bibinfo {author} {\bibfnamefont {N.}~\bibnamefont {Piro}}, \bibinfo {author}
  {\bibfnamefont {R.}~\bibnamefont {Schilling}}, \bibinfo {author}
  {\bibfnamefont {A.}~\bibnamefont {Ghadimi}}, \ and\ \bibinfo {author}
  {\bibfnamefont {T.~J.}\ \bibnamefont {Kippenberg}},\ }\href {\doibase
  10.1038/nature14672} {\bibfield  {journal} {\bibinfo  {journal} {Nature}\
  }\textbf {\bibinfo {volume} {524}},\ \bibinfo {pages} {325} (\bibinfo {year}
  {2015})}\BibitemShut {NoStop}%
\bibitem [{\citenamefont {Tagantsev}\ \emph {et~al.}(2018)\citenamefont
  {Tagantsev}, \citenamefont {Sokolov},\ and\ \citenamefont
  {Polzik}}]{tagantsev_dissipative_2018}%
  \BibitemOpen
  \bibfield  {author} {\bibinfo {author} {\bibfnamefont {A.~K.}\ \bibnamefont
  {Tagantsev}}, \bibinfo {author} {\bibfnamefont {I.~V.}\ \bibnamefont
  {Sokolov}}, \ and\ \bibinfo {author} {\bibfnamefont {E.~S.}\ \bibnamefont
  {Polzik}},\ }\href {\doibase 10.1103/PhysRevA.97.063820} {\bibfield
  {journal} {\bibinfo  {journal} {Physical Review A}\ }\textbf {\bibinfo
  {volume} {97}},\ \bibinfo {pages} {063820} (\bibinfo {year}
  {2018})}\BibitemShut {NoStop}%
\end{thebibliography}%

\end{document}